\documentclass[12pt]{article}
\begin{document}
\setcounter{page}{0}
\thispagestyle{empty}
\setlength{\parindent}{1.0em}
\begin{flushright}
GUTPA/04/06/02
\end{flushright}
\renewcommand{\thefootnote}{\fnsymbol{footnote}}
\begin{center}{\LARGE{{\bf Hierarchy-Problem and a Bound State of 
6 \lowercase{t} and 6 $\overline{\lowercase{t}}$\footnote{\it We dedicate 
this article to Paul Frampton, our friend for many years, on the occasion 
of his sixtieth birthday. It is to be published in the Proceedings of the 
Coral Gables Conference on High Energy Physics and Cosmology, 
Fort Lauderdale, Florida, 17 - 21 December 2003.} }}}
\end{center}
\begin{center}{\large{C. D. Froggatt}\\}
\end{center}
\begin{center}{{\it Department of Physics and Astronomy}\\{\it
University of Glasgow, Glasgow G12 8QQ, Scotland}}\end{center}
\begin{center}{\large{H. B. Nielsen}\\}
\end{center}
\begin{center}{{\it  Niels Bohr Institute, }\\{\it
Blegdamsvej 17-21, DK 2100 Copenhagen, Denmark}}
\end{center}
\begin{center}{\large{L. V. Laperashvili}\\}
\end{center}
\begin{center}{{\it  Institute of Theoretical and Experimental Physics, }\\
{\it Cheremushkinskaya ulica 25, Moscow}}\end{center}
\renewcommand{\thefootnote}{\arabic{footnote}}
\setcounter{footnote}{0}

\begin{center}
{\bf Abstract}
\end{center}

{\small We propose a unification of some fine-tuning problems 
-- really in this article only the problem of why the weak scale is so 
small in energy compared to a presumed fundamental scale, being say the Planck 
scale -- by postulating the zero or very small value of the 
cosmological constant not only for one but for several vacua. 
This postulate corresponds to what we have called the Multiple 
Point Principle, namely that there be many ``vacuum'' states 
with the same energy density. We further assume that 6 top quarks and 
6 anti-top quarks can bind by Higgs exchange so strongly as to 
become tachyonic and form a condensate. This gives rise to the 
possibility of having a phase transition between vacua with and without 
such a condensate. The two vacua distinguished by such a condensate 
will have the same cosmological constant provided the top Yukawa coupling 
is about $1.1 \pm 0.2$, in good correspondence with the experimental value. 
The further requirement that this value of the Yukawa coupling, at the 
weak scale, be compatible with the existence of a third vacuum, with 
a Higgs field expectation value of the order of the fundamental scale, 
enforces a hierarchical scale ratio between the fundamental and weak 
scales of order $10^{16}$ -- $10^{20}$.}

\newpage

\section{Introduction}

The present article has the purpose of suggesting the following two 
relatively new ideas:

1) A unified fine-tuning principle, according to which there exist 
{\em several} vacuum states having zero or approximately zero value for the 
cosmological constant.

2) The existence of a bound state of six top quarks and six anti-top 
quarks whose binding, due to Higgs particle exchange, is so strong 
that a condensate of such bound states could form and make up a phase in 
which essentially tachyonic bound states of this type fill the vacuum. 

There are several fine-tuning problems in the Standard Model (SM): the tiny 
values of the cosmological constant and the strong CP violating parameter 
$\Theta_{QCD}$, the small hierarchy problem of the Yukawa couplings and the 
large hierarchy problem of the weak to fundamental (Planck) scale ratio. 
However these are {\em only} fine-tuning problems and do not 
necessarily require a modification\footnote{Of course neutrino masses 
indicate some new physics at the see-saw scale, which requires a minor 
modification of the SM.} of the SM. They could {\em a priori} be resolved by 
a general fine-tuning principle, which we take in the form of a zero 
cosmological constant postulate combined with our so-called Multiple 
Point Principle \cite{brioni} of degenerate vacua. Thus, much like in 
supersymmetry, our idea is to  postulate that there be many vacuum states 
all having zero or rather approximately zero cosmological constant, but 
without having supersymmetry! We do not speculate here on the underlying 
mechanism responsible for such degenerate vacua, but it seems likely 
that some kind of non-locality is required \cite{brioni}. 
 
In this article we point out the possible existence of at least 3 degenerate 
vacua in the pure SM, which could be responsible for the hierarchy between 
the fundamental and weak scales. In particular we emphasize our second 
idea that one of these vacua is due to the condensation of an exotic 
meson consisting of 6 $t$ and 6 $\overline{t}$ quarks \cite{itepportoroz}. 
The reason that 
such a strongly bound exotic meson has been overlooked until now is 
that its binding is based on the collective effect of attraction between 
several quarks due to Higgs exchange. The effect builds up for many (here 12) 
particles in an analogous way to that of the universal gravitational force of 
attraction, as we now describe.   

\section{Proposed bound state of 6 top quarks and 6 anti-top quarks}
\label{nbs}

As emphasized above, the virtual exchange of the Higgs particle between 
two quarks, two anti-quarks or a quark anti-quark pair yields an attractive 
force in each case. We now consider putting more and more $t$ and 
$\overline{t}$ quarks together in the lowest energy relative S-wave 
states. The Higgs exchange binding energy for the whole system becomes 
proportional to the number of pairs of constituents, rather than to the 
number of constituents. So {\em a priori}, by combining sufficiently many 
constituents, the total binding energy could exceed the constituent 
mass of the system! However we can put a maximum of $6t + 6\overline{t}$ 
quarks into the ground state S-wave. So let us now estimate the binding 
energy of such a 12 particle bound state.

As a first step we consider the binding energy $E_1$ of one of them to 
the remaining 11 constituents treated as just one particle analogous 
to the nucleus in the hydrogen atom. We assume that the radius of the 
system turns out to be reasonably small, compared to the Compton wavelength 
of the Higgs particle, and use the well-known Bohr formula for the binding 
energy of a one-electron atom with atomic number $Z=11$ to obtain the crude 
estimate:
\begin{equation}
E_1 = -\left(\frac{11g_t^2/2}{4\pi}\right)^2 \frac{11m_t}{24}.
\label{binding}
\end{equation}
Here $g_t$ is the top quark Yukawa coupling constant, in a normalisation 
in which the top quark mass is given by $m_t = g_t \, 174$ GeV. 

The non-relativistic binding energy $E_{binding}$ of the 12 particle system 
is then obtained by multiplying by 12 and dividing by 2 to avoid 
double-counting the pairwise binding contributions. This estimate only takes 
account of the $t$-channel exchange of a Higgs particle between the 
constituents. A simple estimate of the $u$-channel Higgs exchange
contribution \cite{itepportoroz} increases the binding energy by a further
factor of $(16/11)^2$, giving:
\begin{equation}
 E_{binding} = \left(\frac{11g_t^4}{\pi^2}\right)m_t
\label{binding2}
\end{equation}

We have so far neglected the attraction due to the exchange of gauge 
particles. So let us estimate the main effect coming from gluon 
exchange\footnote{Note that we here improve our earlier 
estimates \cite{itepportoroz}.} with a QCD fine structure constant 
$\alpha_s(M_Z) = g_s^2(M_Z)/4\pi = 0.118$,
corresponding to an effective gluon $t-\overline{t}$ coupling constant 
squared of:
\begin{equation}
e_{tt}^2 = \frac{4}{3}g_s^2 \simeq \frac{4}{3} 1.5 \simeq 2.0
\end{equation}
For definiteness, consider a $t$ quark in the bound state; it interacts 
with 6 $\overline{t}$ quarks and 5 $t$ quarks. The 6 $\overline{t}$ 
quarks form a colour singlet  and so their combined interaction with the 
considered $t$ quark vanishes. On the other hand the 5 $t$ quarks combine 
to form a colour anti-triplet, which together interact like a $\overline{t}$ 
quark with the considered $t$ quark. So the total gluon interaction of the 
considered $t$ quark is the same as it would have with a single 
$\overline{t}$ quark. In this case the $u$-channel gluon contribution 
should equal that of the $t$-channel. Thus we should compare the 
effective gluon coupling strength $2 \times e_{tt}^2 \simeq 2 \times 2 = 4$
with $(16/11) \times Zg_t^2/2 \simeq 16 \times 1.0/2 = 8$ from the Higgs 
particle. This leads to an increase of $E_{binding}$ by a factor of 
$(\frac{4+8}{8})^2 = (3/2)^2$, giving our final result:
\begin{equation}
 E_{binding} = \left(\frac{99g_t^4}{4\pi^2}\right)m_t
\label{binding3}
\end{equation}

We are now interested in the condition that this bound state should 
become tachyonic, $m_{bound}^2 < 0$, in order that a new vacuum phase 
could appear due to Bose-Einstein condensation. For this purpose we 
consider a Taylor expansion in $g_t^2$ for the mass {\em squared} of the 
bound state, crudely estimated from our non-relativistic binding 
energy formula:
\begin{eqnarray}
m_{bound}^2 & = & \left(12m_t\right)^2 - 2\left(12
m_t\right)\times
E_{binding} + ...\\
& = & \left(12m_t\right)^2\left(1 -\frac{33}{8\pi^2}g_t^4 +
...\right)
\label{expansion}
\end{eqnarray}  
Assuming that this expansion can, to first approximation, be trusted 
even for large $g_t$, the condition $m_{bound}^2=0$ for the appearance 
of the above phase transition with degenerate vacua becomes to 
leading order:
\begin{equation}
\label{gtphase}
g_t|_{phase \ transition} =
\left(\frac{8\pi^2}{33}\right)^{1/4} \simeq 1.24
\end{equation}   

We have of course neglected several effects, such as weak gauge boson 
exchange, $s$-channel Higgs exchange and relativistic corrections. In 
particular quantum fluctuations in the Higgs field could have an important 
effect in reducing $g_t|_{phase \ transition}$ by up to a factor of 
$\sqrt{2}$, as discussed in Section \ref{fluctuations}. It is therefore quite 
conceivable that the value of the top quark running Yukawa coupling constant, 
predicted from our vacuum degeneracy fine-tuning principle, could be in 
agreement with the experimental value 
$g_t(\mu_{weak})_{exp} \approx 0.95 \pm 0.03$. 
Assuming this to be the case, we now make a further application of our 
fine-tuning  principle to postulate the existence of a third degenerate 
vacuum, in which the SM Higgs field has a vacuum expectation value of order 
the fundamental scale $\mu_{fundamental}$.

\section{Three degenerate vacua and the huge 
scale ratio}

In this section, we explain how our degenerate vacuum fine-tuning 
principle can be used to derive the huge ratio, 
$\mu_{fundamental}/\mu_{weak}$, between the fundamental and weak scales.
The basic idea is to use this principle to tune the value of the running 
top quark Yukawa coupling $g_t(\mu)$ {\em both} at the weak scale, as described
above, and at the fundamental scale. Since running couplings vary 
logarithmically with scale, the predicted values $g_t(\mu_{weak})$ and 
$g_t(\mu_{fundamental})$ can easily imply an exponentially large 
scale ratio.

In order to tune the value of $g_t(\mu_{fundamental})$ we postulate the 
existence of a third degenerate vacuum, in which the SM Higgs field has a 
vacuum expectation value of order $\mu_{fundamental}$. For large values of 
the SM Higgs field $\phi \sim \mu_{fundamental} \gg \mu_{weak}$, the 
renormalisation group improved effective potential is well approximated by 
\begin{equation}
V_{eff}(\phi) \simeq \frac{1}{8}\lambda (\mu = |\phi | ) |\phi |^4
\end{equation}
and the degeneracy condition means that $\lambda(\mu_{fundamental})$ 
should vanish to high accuracy. The effective potential $V_{eff}$ 
must also have a minimum and so its derivative should vanish. Therefore  
the vacuum degeneracy requirement means that the Higgs self-coupling 
constant and its beta function should vanish near the fundamental 
scale:
\begin{equation}
\lambda(\mu_{fundamental}) = \beta_{\lambda}(\mu_{fundamental}) = 0
\end{equation}
This leads to the fine-tuning condition \cite{fn2}
\begin{equation}
\label{gt4} 
g_t^4 = \frac{1}{48} \left(9g_2^4 + 6g_2^2g_1^2+3g_1^4 \right)
\end{equation}
relating the top quark Yukawa coupling $g_t(\mu)$ and the electroweak
gauge coupling constants $g_1(\mu)$ and $g_2(\mu)$ at 
$\mu = \mu_{fundamental}$.
We must now input the experimental values of the electroweak gauge 
coupling constants, which we evaluate at the Planck scale using the 
SM renormalisation group equations, and obtain our prediction:
\begin{equation} 
g_t(\mu_{fundamental}) \simeq 0.39.  
\end{equation}
However we note that this value of $g_t(\mu_{fundamental})$, determined 
from the right hand side of Eq.~(\ref{gt4}), is rather insensitive to
the scale, varying by approximately $10\%$ between $\mu = 246$ GeV
and $\mu = 10^{19}$ GeV.

We now estimate the fundamental to weak scale ratio by
using the leading order SM beta function for the top quark 
Yukawa coupling $g_t(\mu)$:
\begin{equation}
 \beta_{g_t} = \frac{dg_t}{d\ln\mu} =
 \frac{g_t}{16\pi^2}\left(\frac{9}{2}g_t^2 - 8g_3^2 -
 \frac{9}{4}g_2^2 - \frac{17}{12}g_1^2\right)
 \label{betatop}
\end{equation}
where the $SU(3) \times SU(2) \times U(1)$ gauge coupling
constants are considered as given. 
It should be noticed that,
due to the relative smallness of the fine structure constants
$\alpha_i =g_i^2/4\pi$ and particularly of
$\alpha_3(\mu_{fundamental})$, the
beta function $\beta_{g_t}$ is numerically rather small at the
fundamental scale. Hence we need many $e$-foldings between the two 
scales, where $g_t(\mu_{fundamental}) \simeq 0.39$ and 
$g_t(\mu_{weak}) \simeq 1.24$. The predicted scale ratio is quite 
sensitive to the input value of $\alpha_3(\mu_{fundamental})$. If we 
input the value of $\alpha_3 \simeq 1/54$ evaluated at the Planck 
scale in the SM, we predict the scale ratio to be
$\mu_{fundamental}/\mu_{weak} \sim 10^{16} - 10^{20}$. 
We note that, as the rate of logarithmic running of $g_t(\mu)$ increases as 
$\alpha_3$ increases, the value of the weak scale is naturally fine-tuned  
to be a few orders of magnitude above the QCD scale. We also 
predict \cite{fn2} the Higgs mass $M_H=135 \pm 9$ GeV. 

\section{Phenomenology of the bound state}

\subsection{Rho parameter}

Strictly speaking, it is {\em a priori} not obvious within our scenario 
in which of the two 
degenerate vacua discussed in Section \ref{nbs} we live. There is 
however one argument that we live in the phase without a condensate of 
new bound states rather than in the one with such a condensate. 
The reason is that such a condensate is not invariant under the 
$SU(2) \times U(1)$ electroweak gauge group and would contribute to 
the squared masses of the $W^{\pm}$ and $Z^0$ gauge bosons. Although 
these contributions are somewhat difficult to calculate, preliminary 
calculations indicate it is 
unlikely that, by some mathematical accident, they should be in the 
same ratio as those from the SM Higgs field. This ratio is essentially 
the $\rho$-parameter, which has been measured to be in accurate agreement 
with the SM value without a new bound state condensate. So we conclude 
that we live in a phase without a condensate of new bound states.

We have previously \cite{itepportoroz} given a weak argument in favour 
of the phase with a condensate emerging in the early Universe out of 
the Big Bang. However, even if it were valid, one could imagine that 
a phase transition occurred, in our part of the Universe, from a 
metastable phase with a bound state condensate into the present one 
without a condensate. A phenomenological signal for such a phase transition 
would be the slight variation of various coupling constants, on a very 
large scale, from region to region in cosmological space and time.   

\subsection{Seeing a bound state of 6$t$ + 5 $\overline{t}$ ?}

We expect the new bound state to be strongly bound and very long lived 
in our vacuum; it could only decay into a channel in which all 12 
constituents disappeared together. The production cross-section of such 
a particle would also be expected to be very low, if it were just crudely 
related to the cross section for producing 6 $t$ and 6 $\overline{t}$ 
quarks. It would be weakly interacting and difficult to detect. There would 
be a better chance of observing an effect, if we optimistically assume that 
the mass of the bound state is close to zero (i.e.~very light compared to 
$12m_t \approx 2$ TeV and possibly a dark matter candidate) even in the 
phase in which we live. In this case 
the bound state obtained by removing one of the 12 quarks would also be 
expected to be light. These bound states with radii of order $1/m_t$ 
might then be smaller than or similar in size to their Compton 
wavelengths and so be well described by effective scalar and Dirac fields 
respectively. The 6 $t$ + 6 $\overline{t}$ bound state would couple 
only weakly to gluons whereas the 6$t$ + 5 $\overline{t}$ bound state 
would be a colour triplet and be produced like a fourth generation 
top quark at the LHC. If these 11 constituent bound states were pair 
produced, they would presumably decay into the lighter (undetected) 
12 constituent bound states with the emission of a $t$ and a 
$\overline{t}$ quark.  

\subsection{Fine-tuning the top mass; Higgs field fluctuations}
\label{fluctuations}

The crucial phenomenological test of our fine-tuning principle is of 
course that it correctly predicts the experimental values of physical 
parameters. The predicted existence of a new phase at the weak scale 
and the value of the top Yukawa coupling $g_t|_{phase \ transition}$ 
at the phase transition provide, in principle, a very clean test, since 
it only involves SM physics. However, in practice, the calculation of the 
binding energy of the proposed 6 $t$ + 6 $\overline{t}$ bound state is 
hard and indeed Eq.~(\ref{gtphase}) overestimates $g_t$. So here we 
consider a potentially large correction due to quantum fluctuations in the 
Higgs field.

The fluctuations in the average of the Higgs field over the interior of 
the bound state get bigger and bigger, as the top Yukawa coupling is 
increased and the size of the bound state diminishes. There is then a 
significant chance that the average value would turn out to be negative 
compared to the usual vacuum value. By thinking of the top quark Dirac 
sea configuration in the bound state, we see that for a sign-inverted 
Higgs field this configuration becomes just the {\em vacuum} state. Such 
a sign-inverted configuration may perhaps best be described by saying 
that neither the non-relativistic kinetic term for the quarks nor their 
mass energy are present, both being in these situations approximated 
by zero. Let us denote by $P_v$ the probability of fluctuating into such 
a vacuum configuration. The most primitive way to take the effect of 
these fluctuations into account is to correct the constituent mass 
in the bound state from $m_t$ to $(1-P_v) m_t$, and the non-relativistic 
kinetic term for the same constituents from $\vec{p}^2/(2m_t)$ to 
$(1-P_v)\vec{p}^2/(2m_t)$. 

It is the kinetic term which determines the binding energy and the 
above correction corresponds to increasing $m_t$ by a factor $1/(1-P_v)$ 
in the binding energy. Therefore the binding energy, which for dimensional 
reasons is proportional to the $m_t$ occurring in the kinetic term (for 
fixed $g_t$), will increase by this factor $1/(1-P_v)$. On the other hand
the constituent masses are corrected the opposite way, meaning that they 
decrease from $m_t$ to $ (1-P_v)m_t$. So the ratio of the binding energy to
the constituent energy -- the binding fraction one could say -- 
increases by the square of the factor $1/(1-P_v)$.

In principle we should now calculate the probability $P_v$ of a 
sign fluctuation as a function of $g_t$. The probability $P_v$ is 
expected to increase as a function of $g_t$ for two reasons: the 
reduction of the Higgs field inside the bound state and its 
decreasing radius. We note that both these effects are more important 
for a bound state of 12 constituents than for, say, toponium. We can, 
however, not expect the fluctuation probability to go beyond 
$P_v = 1/2$. So, for a crude orientation, let us calculate the 
correction in this limiting case. In this case the ratio of the 
binding energy to the constituent energy, which is proportional to 
$g_t^4$, should be increased by the factor $(\frac{1}{1-P_v})^2 = 4$. 
Applying this correction to Eq.~(\ref{gtphase}), we obtain the limiting 
value $g_t|_{phase \ transition} \simeq 1.24 /4^{1/4} = 0.88$. 
This value corresponds to the largest possible correction from 
fluctuations and so we take:
\begin{equation}
 g_t|_{phase \ transition} = 1.06 \pm 0.18
\end{equation} 
as our best estimate, which is in good agreement with the experimental value 
$g_t(\mu_{weak})_{exp} \simeq 0.95$ determined from the physical top quark 
mass.

\end{document}